\documentclass[final,1p,times]{elsarticle} 
\usepackage{graphicx} 
\usepackage{amssymb} 
\usepackage{amsthm} 
\usepackage{lineno} 

\journal{Nuclear Physics A} 
\begin{document} 

\begin{frontmatter} 


\title{The Compressed Baryonic Matter Experiment at FAIR:\\ Progress with feasibility studies and detector developments}

\author{Johann M. Heuser for the CBM collaboration}

\address{GSI Helmholtz Center for Heavy Ion Research GmbH, 
Planck Str. 1, 64291 Darmstadt, Germany}

\begin{abstract} 
The Compressed Baryonic Matter (CBM) experiment is being planned at the international research center FAIR, under realization next to the GSI laboratory in Darmstadt, Germany. Its physics programme addresses the QCD phase diagram in the region of highest
net baryon densities. Of particular interest are the expected first order phase
transition from partonic to hadronic matter, ending in a critical point, and
modifications of hadron properties in the dense medium as a signal of chiral
symmetry restoration. Laid out as a fixed-target experiment at the heavy-ion synchrotrons SIS-100/300, the detector will record both proton-nucleus and nucleus-nucleus collisions at beam energies between 10 and 45$A$ GeV. Hadronic, leptonic and photonic observables have to be measured with large acceptance. The interaction rates will reach 10 MHz to measure extremely rare probes like charm near threshold. Two versions of the experiment are being studied, optimized for either electron-hadron or muon identification, combined with silicon detector based charged-particle tracking and micro-vertex detection. \\
The CBM physics requires the development of novel detector systems, trigger and
data acquisition concepts as well as innovative real-time reconstruction
techniques. Progress with feasibility studies of the CBM experiment and the
development of its detector systems are reported.

\end{abstract} 

\end{frontmatter} 



\section{The CBM physics programme}
The CBM experiment will conduct a comprehensive research programme on nucleus-nucleus collisions at FAIR~\cite{CBM,FAIR}. 
The aim of the CBM project is to investigate the largely unexplored QCD phase diagram at highest net baryon densities and moderate temperatures, complemetary to the heavy ion programmes at RHIC and at LHC that address the physics of the early universe at low densities and high temperatures. 
With projectile energies between 10 and 45$A$~GeV SIS-300 will provide intense beams in the range where the highest net baryon densities are predicted to reach about 10~times that of ground state nuclear matter. 
This will enable the CBM programme to focus on signatures of the expected first order phase transition from partonic to hadronic matter, ending in a critical point, and on modifications of hadron properties, e.g. their masses, in the dense nuclear medium as a signal of chiral symmetry restoration. 
The research programme may already be started at SIS-100 with ion beams between 2 and 11$A$~GeV, and protons up to 29~GeV energy using the HADES detector and an initial version of the CBM experiment. 

Recently proposed other programmes in this beam energy range at RHIC/BNL and NICA/ JINR will be complementary to the CBM programme as they are limited in interaction rates and will focus on bulk particle production. The full exploration including rare probes will be the task of the CBM experiment at FAIR.

\section{The CBM detector}
Two configurations of the CBM detector are being evaluated for electron-hadron and muon-hadron measurements. Both may be realized at different stages. 
They have in common a low-mass silicon tracking system (STS), the central detector to perform charged-particle tracking and high-resolution momentum measurement with radiation tolerant silicon microstrip and pixel detectors.  Combined with an ultra-thin micro-vertex detector (MVD) based on monolithic active pixel sensors,
it will be installed in the gap of a dipole magnet in short distance downstream of the target, typically a gold foil of 250~$\mu$m thickness corresponding to 1\%~nuclear interaction length. 
In the electron-hadron configuration, the CBM experiment comprises a ring imaging Cheren\-kov (RICH) detector downstream of the magnet to identify electron pairs from vector meson decays. Transition radiation detectors (TRDs) provide charged particle tracking and the identification of high energy electrons. Hadron identification will be realized in a time-of-flight (TOF) system built from resistive plate chambers (RPC). An electromagnetic calorimeter (ECAL) will be used to detect direct photons.  The projectile spectator detector (PSD) is a calorimeter that determines centrality and reaction plane of the collisions. 
In the muon-hadron configuration of the experiment, the RICH detector system is replaced by a compact active absorber system (MUCH). Vector mesons are detected via their decays into muon pairs. Hadrons can be measured with the absorbers moved out. 
A particular feature of the experiment is its data acquisition and trigger concept, imposed by the physics programme with rare probes, e.g. charm production near threshold, and the necessity for interaction rates between 0.1 and 10~MHz. 
It is based exclusively on self-triggering front-end electronics that time-stamps the detector signals and ships them to a fast computing farm for event building and high-level trigger generation.

\section{Physics performance studies and detector R\&D}
Progress with the preparation of the experiment has been achieved with detailed physics performance simulations, based on increasingly realistic implementations of the CBM detector systems. This includes feed-back from the beginning detector developments and the evaluation of first demonstrator systems in test-beam experiments~\cite{CBMREPORT}.

\subsection{Charged particle tracking}
The high multiplicity of charged particles is reconstructed with the Silicon Tracking System~\cite{SI-at-FAIR,SI-POSTER}. The measurement of essentially all CBM observables depends on the high performance of the STS. Currently 8 tracking stations are considered in a 1~T dipole magnetic field based on radiation tolerant silicon micro-strip sensors mounted on a light-weight carbon fibre support structure. The material budget may be less then 1\% radiation length per station.  In central 25$A$~GeV Au+Au events, generated with UrQMD, tracks pointing to the primary vertex are reconstructed with 96\% efficiency above momenta of 1~GeV/c using Cellular Automaton and Kalman Filter algorithms. The material budget is such that a momentum resolution of 1.3\% is obtained.  
		
\subsection{Di-lepton spectroscopy}
Electron identification is performed using a Ring Imaging Cherenkov Detector and a Transition Radiation Detector system. The RICH system uses a CO$_2$ radiator of 1.5~m length, providing a pion separation threshold momentum of 4.65~GeV/c. Mirrors of  12~m$^2$ area project electron ring images of about 6~cm diameter onto the 2.4~m$^2$ photo detection plane. The ring finding efficiency, evaluated in simulations with several reconstruction algorithms, is in excess of 95\% for electrons embedded into central Au+Au collisions at 25$A$~GeV beam energy. The detector R\&D for the RICH system focusses on the evaluation of thin spherical glass mirrors of 3~m radius with Al+MgF$_2$ coating and multi-anode photo multiplier tubes coupled to self-triggering readout electronics. For the TRD detector, innovative multi-wire proportional chambers with double-sided pad readout coupled to a foil radiator are being developed, capable of high counting rates.   
The combined electron identification efficiency of RICH and TRD is 85\% at a $\pi$ suppression factor of 10$^4$. The remaining background is dominated by $\pi^0$ Dalitz decays~\cite{DILEPTON}. 

Muon identification is performed with a segmented hadron absorber and a tracking system downstream of the STS. The MUCH system is divided into two regions addressing low-mass muon pairs (vector mesons) and high-mass pairs (J/$\psi$).  
Five iron discs of 20~cm to 35~cm thickness, with a total thickness of 7.5 nuclear interaction lengths, are interleaved with 5~gaps of GEM detector layers and serve the measurement of the low-mass vector mesons. After another 1~m of iron and total 13.5~nuclear interaction lengths, the J/$\psi$ pair candidates are detected. The particle multiplicity is reduced such that after 1.25~m iron 0.25 identified muons are obtained in 25$A$~GeV Au+Au with the dominant background coming from $\pi$ and K decays at 0.13~muons per event~\cite{DILEPTON}. 

\begin{figure}[h]
\centering
\hspace*{-3mm}
\begin{tabular}[b]{cccc}
  \includegraphics[width=0.225\textwidth]{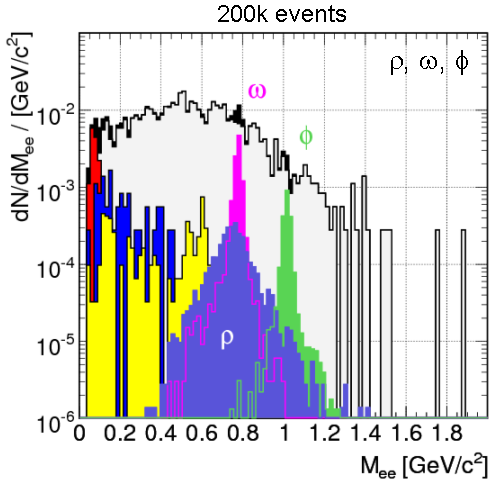} & 
  \includegraphics[width=0.225\textwidth]{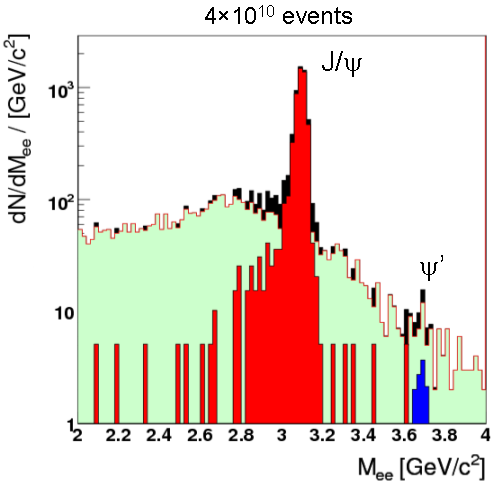} &
  \includegraphics[width=0.225\textwidth]{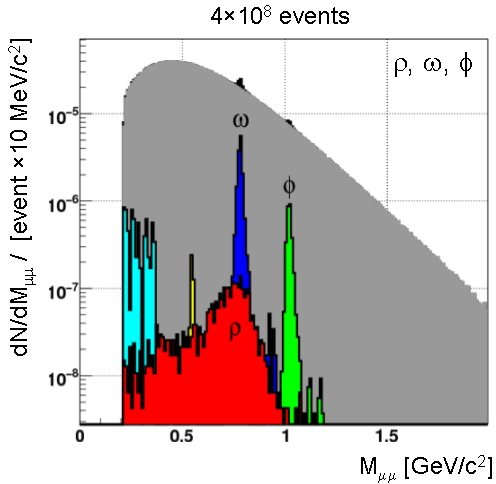} &
  \includegraphics[width=0.225\textwidth]{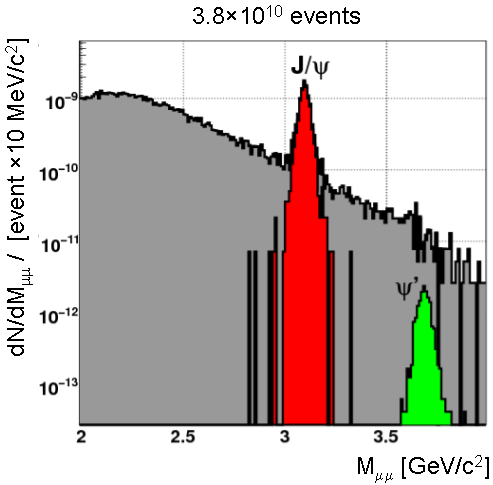} \\
\end{tabular}
\caption[]{Di-lepton invariant mass spectra (signal and combinatorial background) in the low and high-mass regions for minimum-bias Au+Au events at 25$A$~GeV/nucleon. Left two plots: Di-electron signal; Right two plots: Di-muon signal. }
\vspace*{-2mm}
\label{DILEPTON-MASS}
\end{figure}

\begin{figure}[h]
\centering
\hspace*{-3mm}
\begin{tabular}[b]{cccc}
  \includegraphics[width=0.24\textwidth]{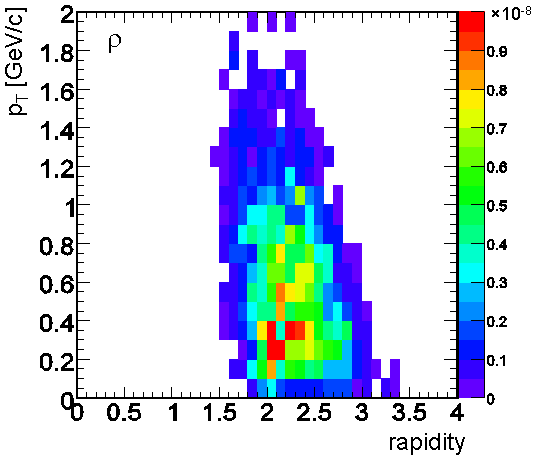} & 
  \includegraphics[width=0.23\textwidth]{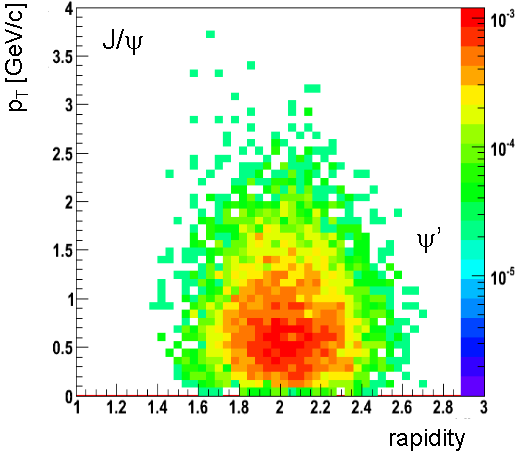} &
  \includegraphics[width=0.221\textwidth]{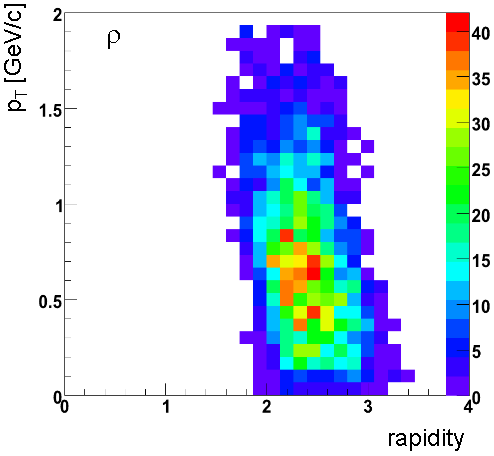} &
  \includegraphics[width=0.227\textwidth]{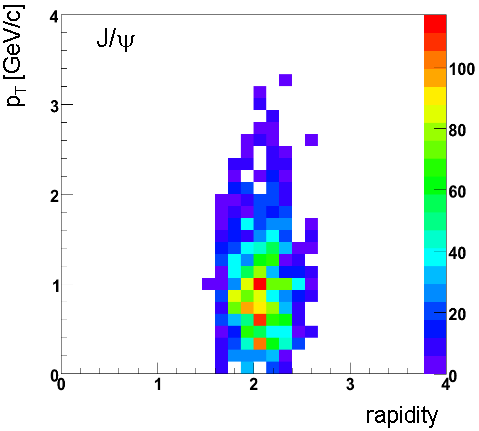} \\
\end{tabular}
\caption[]{Phase space of reconstructed di-leptons. Left two plots: Di-electron signal; Right two plots: Di-muon signal.}
\vspace*{-2mm}
\label{DILEPTON-PHASE}
\end{figure}

Spectra and phase-space coverage of reconstructed di-leptons are shown in Figs.~\ref{DILEPTON-MASS} and ~\ref{DILEPTON-PHASE}.

\subsection{Hadron measurement}
Hadrons are identified via time-of-flight measurement with a detector system based on resistive plate chambers. Different RPC technologies are being explored, including ultra-thin glass pad RPC, ceramic RPC, and differential RPC based on semi-conductive glass. High counting rate capability up to 25~kHz has to be achieved with a time resolution of 80~ps~\cite{RPC}. 
The ``global'' track reconstruction efficiency obtained with the simulated STS, TRD and TOF systems is 85\%.

\subsection{Open charm detection}
\begin{figure}[h]
\centering
\vspace*{2mm}
\includegraphics[width=0.7\textwidth]{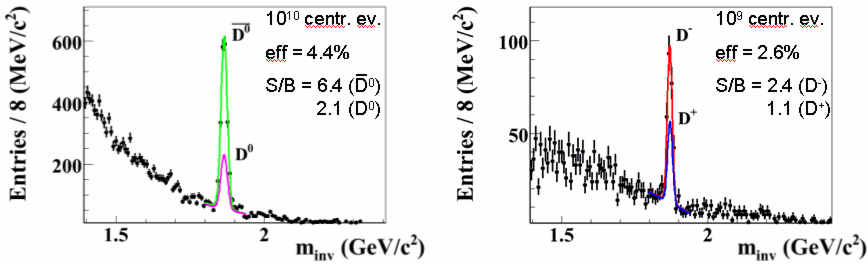}
\vspace*{-2mm}
\caption[]{Invariant mass spectra of D mesons in central Au+Au collisions of 25$A$~GeV. }
\label{DMASS}
\vspace*{-4mm}
\end{figure}

The efficient separation of primary and short-lived decay vertices for open charm detection requires a z-vertex resolution of the order of 50~$\mu$m r.m.s. This is planned to be realized with a two-station micro vertex detector in front of the silicon tracker, built from Monolithic Active Pixel Sensors~\cite{MVD-CBM}. The stations must be ultra-thin. Two MVD stations with a material budget of 0.3\% and 0.5\% radiation lengths have been studied at 5~cm and 10~cm downstream of the target. No K and p identification is performed but proton rejection via TOF.  The interaction rate is limited to 10$^5$/s as imposed by the maximum readout speed of the monolithic pixel detectors. With this MVD system, a data sample of about 7k~D$^0$ + 20k~$\bar{\mbox{D}}$$^0$ and 12k~D$^+$ + 26k~D$^-$ will be collected in 10~weeks of data taking with efficiencies as shown in Fig.~\ref{DMASS}.  

\subsection{Data acquisition and on-line event selection}
Further developments comprise self-triggering front-end electronics for the STS, RICH and MUCH systems, a readout chip for the TOF system with 25~ps time resolution, a data acquisition system with 500~MB/s/node throughput, and fast on-line event selection and track reconstruction. For the on-line computing many-core architectures are explored (GPUs, CELL, LRB)~\cite{CBM-tracking}. A~starting point is the application of commercial high-end graphics cards assembled into a GPU farm. 
The maximum archiving rate will be 25 kHz. High-level trigger strategies are being developed for different physics cases, including: (a)~Open charm: full event reconstruction;	limited by MVD to 10$^5$-10$^6$ events/s; 
(b)~J/$\psi$,$\omega$, $\phi$~$\rightarrow$~$\mu^+$$\mu^-$: event pre-selection by MUCH ($\times$10$^{-3}$) and (c)~J/$\psi$~$\rightarrow$~e$^+$e$^-$:	trigger based on TRD and STS (minimum bias events).


\section*{Acknowledgments} 
The work reported on received support from the EU Integrated Infrastructure Initiative Ha\-dronPhysics, HadronPhysics2, INTAS , ISTC, the Bundesministerium f\"ur Bildung und Forschung, the GSI~Hochschulprogramm MAMAEN, and ROSATOM.

\end{document}